\documentclass[aps,prl,showpacs,twocolumn]{revtex4}
\usepackage{graphicx}
\usepackage{amsmath}
\usepackage{amstext}
\usepackage{latexsym}
\usepackage{amsfonts}
\usepackage{bm}
\usepackage{amssymb}
\usepackage{amscd}
\begin{document}
\title{Non-deterministic Two-Way Quantum Key Distribution using Coherent States}
\author{Won-Ho Kye}
\email{whkyes@empal.com}
\affiliation{The Korean Intellectual Property Office, Daejeon
302-701, Korea}

\date{\today}
\begin{abstract}
We propose a non-deterministic two-way quantum key distribution in which the quantum correlation is established by transmitting the randomly polarized photon.
We analyze the security of the proposed quantum key distribution 
against photon number splitting, impersonation, and Trojan horse attack and quantify the security bound against
mean photon number of the coherent state pulse.
Finally, we remark the characteristic features of the protocol.

\end{abstract}
 \maketitle
Quantum key distribution (QKD) \cite{BB84, E91, Gisin} is to 
generate shared secret information between distant parties
with negligible leakage of the information to an eavesdropper Eve.
The security of QKD is based on the no-cloning theorem: Eve
can not extract any information without introducing errors \cite{Gisin},
while the security of the classical key distribution or cryptography is supported 
by the computational complexity of the underling mathematical problems \cite{RSA}.
Since the QKD by Bennent and Brassard (BB84) \cite{BB84}, 
there has been security proof \cite{Proof} and theoretical proposals to enhance the security \cite{Decoy}. 

The Ping-Pong protocol (PP) proposed by Bostr\"om et al. \cite{Ping} is the first two-way quantum key
distribution based on entangled qubit. It is a conceptually new scheme in the sense 
that the key is generated by the round trip of the qubit, while in the conventional 
QKD it is done with single trip. With this trend, Lucamarini et al. \cite{Luca} 
proposed the two-way protocol without entanglement by merging the
peculiarities of BB84 and PP and recently, Kye et. al., proposed to 
the three-way QKD \cite{Kye} 
to make the encoding possible with relatively dense coherent-state pulse.
One of the interesting aspects of their protocol is to use 
the qubit with random polarization, while in the conventional protocol, predefined finite number 
of polarization states is used \cite{BB84, E91, Gisin, Ping, Luca}. 

In the multi-way quantum key distribution \cite{Ping, Luca, Kye}, 
the fact that the final key is led 
by only one party which allows creating the key in deterministic way is considered as an 
advantage for the direct encoding. However in some cases the deterministic feature of QKD 
without dissipation of the qubit often provides Eve 
with the room to track the protocol easily \cite{Ping-Attack, Kye-Attack}.
That is to say the deterministic feature can play a role of potential security hole in QKD. 

In this paper, we propose a non-deterministic two-way QKD in which the quantum correlation
is established by transmitting the randomly polarized photon.
The initial random polarization  $|\theta\rangle$ with arbitrary $\theta \in [0, \pi]$ is 
compensated by acting the unitary operator $U(-\theta)$ on the returning qubit and the net 
encoding  information can be extracted from that. 
In addition, the non-deterministic feature comes from the $N$ number of screening angle which is chosen by Alice
at the initial stage of the protocol. 
The key is created only when the matching condition of the corresponding screening angles is satisfied 
and it plays important role in blocking up the impersonation and
Trojan horse attack.

QKD using coherent-state pulse has received much attentions in regard to the
practical implementation \cite{Gisin, Gisin1,Kye}.
Since there is no phase reference outside Alice or Bob's lab, 
a coherent-state $|\sqrt{\mu} e^{i \theta}\rangle $ 
of mean photon number $\mu$ is effectively described by
photon number eigenstate $|n\rangle$ with Poission distribution:
$|\sqrt{\mu}\rangle=\exp(-\mu/2)\sqrt{\mu}^n/\sqrt{n!} |n\rangle$ \cite{Coherent}.
As we shall see, our proposal has remarkable advantages for implementation 
using coherent-state pulse, because the protocol allows 
not only to transmit the relative dense coherent pulse 
but also to increase the raw key creation rate.  
Our protocol is described as follows:
\\

{\it Protocol:}

\begin{enumerate}

\item[(P.1)] Alice and Bob initiate the protocol by announcing a set $S(N)$ which has $N$ number of screening angles,  
\begin{eqnarray}
 S(N)= \{ \alpha_1, \cdots \alpha_N \},
\end{eqnarray}
where $N \geq 2$ and the screening angle $\alpha_i$ is defined as $\alpha_i=i\pi/(N+1)$.
 
\item[(P.2)] Alice
take arbitrary angle $\theta$ and chooses screening angle $\alpha_a$ 
and random screening factor $s\in\{0, 1\}$.
She prepares the qubit:
\begin{equation}
	|\theta + \delta_{0s}\alpha_a\rangle,
\end{equation}
where $\delta_{pq}=1$ for $p=q$ otherwise $\delta_{pq}=0$.
Alice occasionally takes $\theta$ with the probability $c$ as a predefined value 
$\theta^* \in \{0, \pi/2\}$ which is called authentication angle.
If $\theta=\theta^*$
then the protocol follows the authentication mode (A-Mode)
else transmission mode (T-Mode).

\begin{enumerate}

\item [(A-mode 1)]Bob chooses a screening angle $\alpha_b$. 
He acts $U((-1)^k\pi/4+\alpha_b)$ on the received qubit, where $k$ is key bit.
The qubit becomes
\begin{equation}
|\theta^*+(-1)^k \pi/4 +\delta_{0s}\alpha_a+\alpha_b\rangle,
\end{equation}
The fraction $(1-t)$ of the 
photons in the qubit enter into the Bob's detector, where $t$ is the 
transmission efficiency of Bob's detector.
Bob records the outcome $O_b$ in his detector.

\item [(A-mode 2)]
After acting $U(-\theta^*+\delta_{1s}\alpha_a)$ on the returning qubit, 
Alice has the qubit $|(-1)^k \pi/4 +\alpha_b+\alpha_b\rangle$.
She measures the qubit and the outcome is recorded as $O_a$. 

\item [(A-mode 3)]
Alice declares the mode is A-mode, then Alice and Bob announce
the chosen screening angles $\alpha_a$ and $\alpha_b$, respectively.
If the screening angles satisfy that
\begin{equation}
\alpha_a+\alpha_b=\pi,
\label{eq.sum}
\end{equation}
then the qubit incoming to Bob's detector is $|\theta^* + (-1)^k \pi/4 \rangle$.
So the corresponding outcome $O_b$ and encoded key $k$ are correlated by 
\begin{equation}
O_b=k\oplus(2\theta^*/\pi),
\label{eq.ob}
\end{equation}
where $\oplus$ is the addition on $\mod 2$ space.
If the verification is failed, Alice and Bob immediately terminate the protocol and
initiate the protocol form (P.1) later.
If $\alpha_a+\alpha_b\neq\pi$ in above step Alice and Bob return (P.2).
\end{enumerate}

\begin{enumerate}
\item [(T-mode 1)]
Bob chooses a screening angle $\alpha_b$. 
He acts $U((-1)^k\pi/4+\alpha_b)$ on the received qubit, where $k$ is key bit.
The qubit becomes
\begin{equation}
|\theta+(-1)^k \pi/4 +\delta_{0s}\alpha_a+\alpha_b\rangle.
\end{equation}
Bob returns the qubit to Alice.

\item [(T-mode 2)]
Alice acts $U(-\theta+\delta_{1s}\alpha_a)$ on the returning qubit and
it becomes $|(-1)^k \pi/4 + \alpha_a+\alpha_b\rangle$. 
Alice measures the qubit and gets the outcome $O_a$. 

\item [(T-mode 3)]
Alice and Bob announce
the chosen screening angles $\alpha_a$ and $\alpha_b$  respectively.
If the screening angles satisfy that $\alpha_a+\alpha_b=\pi$,
then Alice gets the key $k$ for the outcome $O_a$:
\begin{equation}
O_a=k,
\label{eq.ob}
\end{equation}
else Alice and Bob go to (P.2).
If the desired key length is created then go to (P.3)
\end{enumerate}
\item [(P.3)]
Alice and Bob create key $k_a$ and $k_b$ by concatenating key bits and 
exchange the hash values $h(k_a)$ and $h(k_b)$ \cite{Kye}.
If $h(k_a)=h(k_b)$ then the key creation is finished 
else Alice and Bob start again from (P.1). 
\end{enumerate}

Eq. (5) shows Alice's integrity condition observed in Bob's detector(D0, D1 in Fig. 1), which plays an important
role to detect lethal strategy like impersonation and Trojan Horse attack.
Even though the key encoding is performed by Bob deterministically, 
the final key is rearranged whether the matching condition in Eq. (4) is satisfied or not. 
In the QKD the raw key creation rate depends on the number of screening 
angle and mode probability as
\begin{equation}
R_{raw} = q \mu f_{rep} t_{link}\eta_{det},
\end{equation}
where $q$ depends on implementation ($q=(1-c)/N$ for our protocol), $f_{rep}$ is pulse rate, $t_{link}$ the transmission and $\eta_{det}$ the detection efficiency \cite{Gisin}.
Now, we shall analyze the security of the protocol.\\

\begin{figure}
\rotatebox[origin=c]{0}{\includegraphics[width=7.5cm]{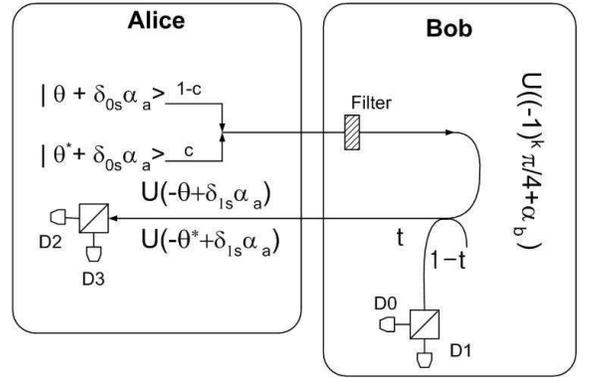}}
\caption{Schematic diagram for the experimental setup. D0 and D1: Bob's detectors; 
D3 and D4: Alice's detectors; PBS: Polarization Beam Splitter. 
Bob has equipped the optical filter to reject undesired frequency. }
\label{figure}
\end{figure}

{\it{Security against photon number splitting (PNS) attack:}}\\
Since A-mode is only for authenticating purpose, 
it is enough to consider that Eve's attack is focused on the T-mode 
in quantifying the PNS attack \cite{PNS}.
As usual, we assume that Eve is so superior that
her action is limited only by the law of physics. 
Against the coherent state $|\sqrt{\mu}\rangle$ from Alice, she replaces
the lossy channel by a perfect one and puts a beam splitter of transmission efficiency $\eta$ in the middle \cite{PNS}.
The reflected field, which is a coherent state with its amplitude $|\sqrt{1-\eta}\sqrt{\mu}\rangle$, will be the
source of information to Eve.  In the protocol (T-mode.1)-(T-mode.3),
the information transmitted between Alice and Bob is of random
polarization.  In our protocol, the photon
polarizations lie on the equator of the Poincar\'e sphere. 
Thus, in this case, Eve's goal is to find the optimum 
state estimation from $n$ qubits gives the maximal
mean fidelity given by \cite{Buzek}:
\begin{equation}
I(n)={1 \over 2}+\frac{1}{2^{n+1}}\sum_{\ell=0}^{n-1}\sqrt{
  \begin{pmatrix} n \\ \ell \end{pmatrix} \begin{pmatrix} n \\ \ell+1
  \end{pmatrix}}.
\label{info-2}
\end{equation}

Let us first consider the maximum information Eve can get from the
Alice$\rightarrow$ Bob channel in (a.2).  The probability of
there being $n$ photons  of the channel in the coherent state
$|\sqrt{(1-\eta)\mu}\rangle$ is 
$P_{AB}(n)=\exp[-(1-\eta)\mu]\frac{[(1-\eta)\mu]^n}{n!}$.
The received qubit in Bob's end is $|\sqrt{\eta \mu}\rangle$  and after transmission of the detector,
it becomes $|\sqrt{\eta t \mu}\rangle$.
Thus in Bob $\rightarrow$ Alice channel, the probability of there being $n$ photons
in the coherent state $|\sqrt{1-\eta}\sqrt{\eta t \mu}\rangle$ is
$P_{BA}(n)=\exp[-(1-\eta)\eta t\mu]\frac{[(1-\eta)\eta t\mu]^n}{n!}$

Then the maximum amount of information Eve can
get from the channel in A$\rightarrow$B and B$\rightarrow$A
is $I_{AB}= \sum_{n=0}^\infty P_{AB}(n)I(n)$ and $I_{BA}=\sum_{n=0}^\infty P_{BA}(n)I(n)$, respectively.    
The maximum information Eve can obtain is bounded by $I_E=\min(I_{AB}, I_{BA})$, which is plotted in Fig. 2 
for various cases.  
Since the intensity of the coherent pulse decreases as the number of laps between Alice and Bob, $I_E$ is
actually determined by $I_{BA}$.

Now we define the critical value of initial amplitude $\alpha^*$ which 
gives the average number of photons delivered to Alice about 1 after (T-mode 3). 
Since the incoming amplitude of the coherent pulse in (T-mode 3) is  $|\sqrt{(1-\eta)\eta t\mu}\rangle$, the
critical value of initial amplitude is given by $\mu^*=1/((1-\eta)\eta t)$. 
At the critical value of initial amplitude, maximum bound for Eve's information $I_E^*=\sum_{n=0}^\infty \frac{\exp(-1)}{n!} I(n) \approx 0.6900$, while the mutual information between Alice and Bob is unity 
(if the detector of Alice is not clicked in a particular time window 
due to the empty pulse, Alice and Bob could exclude the
corresponding event by announcing that the pulse is empty).
That is to say, at the critical amplitude Alice and Bob shares 
$31\%$ higher information than that of Eve. So Alice and Bob could create the 
final key through the post processing like privacy amplification \cite{PA}.\\
We remark the critical mean photon number in our protocol is 
on $ 5 \leq \mu^* \leq 15 $ which is at least ten times
larger value than  $\mu \leq 0.2$ \cite{Exp-mu1, Exp-mu2} of conventional QKD. 
Accordingly, our protocol allow the higher raw key creation rate, even though the
$q$ factor in Eq. (9) is slightly smaller that the conventional QKD.

\begin{figure*}
\rotatebox[origin=c]{0}{\includegraphics[width=15cm]{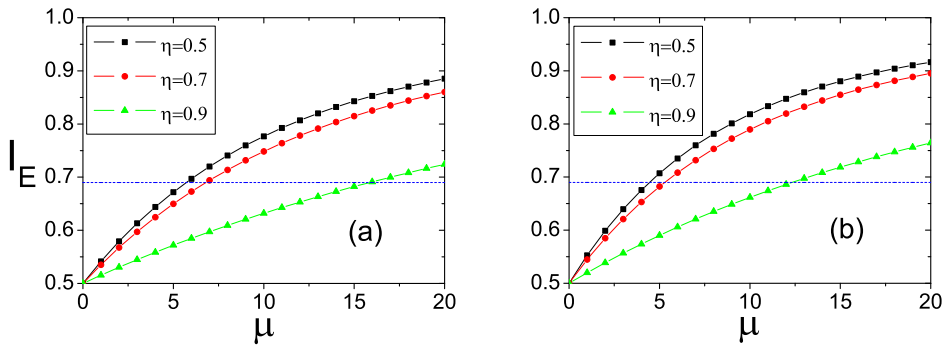}}
\caption{Maximum bound for Eve's information I$_E$ as a function of 
mean photon number $\mu$ of the coherent-state pulse according to various channel transmission efficiency $\eta$ at the
transmission efficiency of Bob's detectors $t=0.7$ (a) and $t=0.9$ (b). The horizontal line 
in (a) and (b) shows the maximum bound of information for Eve 
when Alice prepares the initial amplitude 
as the critical value $\mu=\mu^*$. }
\label{figure2}
\end{figure*}

{\it{Security against Impersonation attack:}}\\
Eve can impersonate Bob to Alice and Bob to Alice in the quantum channel.
This type of attack is effective on the protocol which transmits the qubit 
without dissipation of the qubit \cite{Ping,Kye}.
Against our protocol, Eve may consider the following strategy:
\begin{enumerate}
\item[(A1.1)] After the step (P2) Eve intercepts the qubit and puts it in the quantum storage.
Let's call it $E_1=\{|\theta + \delta_{0s}\alpha_a\rangle \}$.
Eve prepare fake qubit $|\theta^\prime +\delta_{0s^\prime}\alpha_a^\prime \rangle$ and send it to Bob.

\item[(A1.2)] After the step (A-mode 1) or (T-mode1), Eve intercepts again the qubit whose state is given by
$|(-1)^k \pi/4+\theta^\prime +\delta_{0s^\prime}\alpha_a^\prime +\alpha_b\rangle$. Eve gets the qubit
$|(-1)^k \pi/4 + \alpha_b \rangle$ after acting $U(-\theta^\prime -\delta_{0s^\prime}\alpha_a^\prime)$ on the qubit.
Eve measures the qubit with guessing $\alpha_b=\alpha_b^\prime$ and gets the outcome $O_e$ and key $k^\prime = O_e$.

\item[(A1.3)] Eve encodes the intercepted original qubit $E_1$ by 
acting $U((-1)^{k^\prime} \pi/4 +\alpha_b^\prime)$. The intercepted qubit becomes 
$E_1^\prime=\{|(-1)^{k^\prime}\pi/4+\theta + \delta_{0s}\alpha_a+\alpha_b^\prime\rangle \}$. Eve sends the qubit to Alice.
\end{enumerate}
If Eve impersonate the quantum channel during  T-mode,
the probability that Eve's guessing of $\alpha_b$ in (A1.2) was right is $1/N$.
Accordingly Alice's key with Eve's impersonation includes the error with the probability and
it should be noticed in the step (P.3). 

On the other hand, if Eve impersonates the quantum channel during A-mode, 
Eve's impersonation can be also detected at the step (A-mode 3).
In the step (A1.1), Eve's fake qubit $|\theta^\prime +\alpha_a^\prime \rangle$ is 
entered into Bob's detector with the transmission efficiency $(1-t)$. 
In that case, Eve's fake qubit
violates with the integrity condition of Eq. (5) because
the probability that Eve's guessing was matched with Alice's authentication angle as well as screening,
$\theta^\prime =\theta^*$ and $\alpha_a^\prime =\alpha_a$ is almost null (here we use the fact that
Eve does not know if the protocol in A-mode or T-mode). Eve's impersonation should be notice in the step (A-mode 3) with Eq. (5)-(6).\\

{\it{Security against Trojan Horse type attack:}}\\
Eve could attach ancillary qubit to the transmitted qubit and after the Bob's encoding,
she reads out the encoding by measuring the ancillary qubit after separating out the ancillary qubit 
form the unified qubit.
It is shown that the attack strategy is effective on the multiple-way protocol \cite{Ping, Ping-Attack}.
We assume that Alice has a properly designed filter to reject the unnecessary of photons 
with split wave length as in Fig. 1 \cite{Gisin}. So Eve has some difficulty to distinguish ancillary form the full qubit and
eventually she could not separate out the ancillary qubit
form the unified qubit, perfectly. Nevertheless, to demonstrate the robustness of
our protocol, we allow Eve to inject the ancillary qubit which has split wave length compared with
the transmitted qubit and to separate the ancillary qubit from the returning qubit. 

Eve may consider the strategy as follows:
\begin{enumerate}
\item[(A2.1)] After the step (A-mode 1) or (T-mode 1), Eve prepares an ancillary state $|0\rangle$ and 
attaches the ancillary onto the qubit from Alice. The qubit with the ancillary state is
$|\theta + \delta_{0s}\alpha_a\rangle \otimes |0 \rangle$.
Eve sends the qubit to Bob.
\item[(A2.2)]After the step (A-mode 1) or (T-mode 1), the returning qubit becomes
$|(-1)^k \pi/4+\theta + \delta_{0s}\alpha_a+\alpha_b\rangle \otimes |(-1)^k \pi/4 + \alpha_b \rangle$.
Eve separates out ancillary and keep the qubit in storage as $E_2=\{|(-1)^k \pi/4 + \alpha_b \rangle \}$. 
After the step (A-mode 3) or (T-mode 3), Eve knows the Bob's screening angle $\alpha_b$ and measures
the qubit in $A_1$ and reads the key $k$.
\end{enumerate}
If Eve can distinguish if the protocol is in either T-mode or A-mode. 
Eve attacks on the protocol only when the protocol is in T-mode.
In that case, she can read the
key in the fraction of $t$ of the created key, 
because the ancillary qubit sink into the Bob's detector with the fraction of $1-t$.
Unfortunately, there always exists $\theta$ which satisfies $\theta+\delta_{0s}\alpha_a= \theta ^* + \delta_{0s^\prime}\alpha_a^\prime $, 
where $\alpha_a, \alpha_a^\prime \in S(N)$ and $s, s^\prime \in \{0, 1\}$ it inevitably induces the collision between two mode.
Thus Eve can not distinguish the initial qubit state if it is in T-mode or A-mode unambiguously 
and she could not avoid to intervening during A-mode and her ancillary qubit sink into the Bob's detector.
The ancillary qubit which does not carry the information Alice's screening angle and authentication
angle makes the Bob's outcome $O_b^\prime$ violating the integrity condition.
Thus Eve's Trojan Horse attack should be noticed in the step (A-mode 3).

\section{Conclusions}
We have proposed the non-deterministic two-way QKD protocol and have demonstrated the security 
of the proposed protocol against PNS, Impersonation and Torjan Horse attack.
Finally, we emphasize that the proposed protocol has the following advantages compared with the conventional
QKD.
1) The quantum correlation is established by exchanging 
the qubit with completely random polarization. For that reason, 
our protocol can be implemented with relatively dense coherent pulse.
2) Since the mean photon number $\mu$ can be safely set is much 
higher value than that of  conventional QKD \cite{Exp-mu1, Exp-mu2} (see the last paragraph of PNS analysis), 
the corresponding raw key creation rate is higher than that of the conventional two-way QKD.
3) The protocol provides the tunable security depending on the number of screening angle $N$.
Even if an eavesdropper try to know the current status of the protocol
by a combination of photon number quantum non-demolition measurement\cite{PNS}
and unambiguous state discrimination \cite{UD}, it can be avoided 
by increasing the number of screening angle $N$.


\end{document}